\documentclass[12pt]{article}
\usepackage{amsmath}
\usepackage{amssymb}
\usepackage{graphicx}
\usepackage{pstricks}
\usepackage{citesort}
\numberwithin{equation}{section}
\begin{document}

\setcounter{page}{0}
\thispagestyle{empty}

\begin{flushright}
{\small BARI-TH 406/2001}
\end{flushright}

\vspace*{2.5cm}

\begin{center}
{\large\bf Abelian chromomagnetic background field \\[0.4cm]
 at finite temperature on the lattice }
\end{center}

\vspace*{2cm}

\renewcommand{\thefootnote}{\fnsymbol{footnote}}

\begin{center}
{
Paolo Cea$^{1,2,}$\protect\footnote{{\tt Paolo.Cea@bari.infn.it}} and
Leonardo Cosmai$^{2,}$\protect\footnote{{\tt Leonardo.Cosmai@bari.infn.it}} \\[0.5cm]
$^1${\em Dipartimento di Fisica, Universit\`a di Bari,
I-70126 Bari, Italy}\\[0.3cm]
$^2${\em INFN - Sezione di Bari,
I-70126 Bari, Italy}
}
\end{center}

\vspace*{0.5cm}

\begin{center}
{
January, 2001}
\end{center}

\vspace*{1.0cm}

\renewcommand{\abstractname}{\normalsize Abstract}
\begin{abstract}
The vacuum dynamics of SU(2) and SU(3) lattice gauge
theories is studied by means of a gauge-invariant effective action
defined using the lattice Schr\"odinger functional at finite
temperature. In the case of the SU(3) gauge theory numerical
simulations are performed at zero and finite temperature. The
vacuum is probed using an external constant Abelian chromomagnetic
field. At zero temperature, in agreement with our previous studies for the
SU(2) theory, the external field is
totally screened in the continuum limit. 
At finite temperature numerical data
suggest that confinement is restored by increasing
the strength of the applied field.
We give also an estimate of the deconfinement temperature.
\end{abstract}

\vspace*{0.5cm}
\begin{flushleft}
PACS number(s): 11.15.Ha
\end{flushleft}
\renewcommand{\thesection}{\normalsize{\arabic{section}.}}
\section{\normalsize{INTRODUCTION}}
\renewcommand{\thesection}{\arabic{section}}

The lattice approach to gauge theories allows the non-perturbative
study of gauge systems without loosing the gauge invariance.
Recently~\cite{Cea:1997ff} it has been proposed a method to define on
the lattice the gauge invariant effective action by using the
Schr\"odinger
functional~\cite{Rossi:1980pg,Rossi:1980jf,Gross:1981br,Luscher:1992an}.
\\
In the continuum the Euclidean Schr\"odinger functional in
Yang-Mills theories without matter fields reads:
\begin{equation}
\label{Zeta}
{\mathcal{Z}}[A^{(f)},A^{(i)}] = \left\langle
A^{(f)} \left| e^{-HT} {\mathcal{P}} \right| A^{(i)} \right\rangle
\,.
\end{equation}
In Eq.~(\ref{Zeta}) $H$ is the pure gauge Yang-Mills Hamiltonian
in the fixed-time temporal gauge, $T$ is the Euclidean time
extension, while ${\mathcal{P}}$ projects onto the physical
states. $A^{a(i)}_k(\vec{x})$ and $A^{a(f)}_k(\vec{x})$ are static
classical gauge fields, and the state $|A\rangle$ is such that
\begin{equation}
\label{stateA}
\langle A | \Psi \rangle = \Psi[A] \,,
\end{equation}
$\Psi[A]$ being a wavefunctional. Note that, by definition,
${\mathcal{Z}}[A^{(f)},A^{(i)}]$ is invariant under gauge
transformations of the gauge fields $A^{(i)}$ and $A^{(f)}$. \\
Using standard formal manipulations and the gauge invariance of
the Schr\"odinger functional it is easy to rewrite
${\mathcal{Z}}[A^{(f)},A^{(i)}]$ as a functional
integral~\cite{Rossi:1980pg,Rossi:1980jf,Gross:1981br}
\begin{equation}
\label{Zetaint}
{\mathcal{Z}}[A^{(f)},A^{(i)}] = \int {\mathcal{D}}A \;
e^{-\int_0^T dx_4 \, \int d^3x \, {\mathcal{L}}_{YM}(x)}
\end{equation}
with the constraints:
\begin{eqnarray}
\label{constraints}
A_\mu(x_4=0) = & A^{(i)}_\mu \nonumber \\
\\
A_\mu(x_4=T) = & A^{(f)}_\mu \nonumber \; .
\end{eqnarray}
It is worthwhile to stress that in Eq.~(\ref{Zetaint}) we should
sum over the topological inequivalent classes. However, it turns
out~\cite{Luscher:1992an} that on the lattice such an average is not
needed because the functional integral in Eq.~(\ref{Zetaint}) is
already invariant under arbitrary gauge transformation of
$A^{(i)}$ and $A^{(f)}$. The lattice implementation of the
Schr\"odinger functional  Eq.~(\ref{Zetaint}) is given
by~\cite{Luscher:1992an}:
\begin{equation}
\label{Zetalatt}
 {\mathcal{Z}}[U^{f},U^{i}] = \int
{\mathcal{D}}U \; e^{-S} \,,
\end{equation}
where the integrations over the links $U_{\mu}(x)$ are done with
the fixed boundary conditions:
\begin{equation}
\label{boundary}
U(x)|_{x_4=0} = U^{(i)}\,, \quad U(x)|_{x_4=T} = U^{(f)} \,.
\end{equation}
The links $U^{(i)}$ and $U^{(f)}$ are the lattice version of the
continuum gauge fields $A^{(i)}$ and $A^{(f)}$. \\
In Ref.~\cite{Cea:1997ff} we introduced the new functional:
\begin{equation}
\label{Gamma}
\Gamma[\vec{A}^{\mathrm{ext}}] = -\frac{1}{T} \ln \left\{
\frac{{\mathcal{Z}}[U^{\mathrm{ext}}]}{{\mathcal{Z}}(0)} \right\}
\,,
\end{equation}
where
\begin{equation}
\label{ZetaUext}
{\mathcal{Z}}[U^{\mathrm{ext}}] =
{\mathcal{Z}}[U^{\mathrm{ext}},U^{\mathrm{ext}}] \,,
\end{equation}
and ${\mathcal{Z}}[0]$ means the Schr\"odinger functional
Eq.~(\ref{ZetaUext}) without external background field
($U^{\mathrm{ext}}_\mu=\mathbf{1}$). \\
From the previous discussion it is evident that
${\mathcal{Z}}[U^{\mathrm{ext}}]$ is invariant against lattice
gauge transformations of the external link $U^{\mathrm{ext}}$.
Moreover, it can be shown that~\cite{Cea:1997ff}:
\begin{equation}
\label{limit}
\lim_{T \to \infty} \Gamma[\vec{A}^{\mathrm{ext}}] =
E_0[\vec{A}^{\mathrm{ext}}] - E_0[0]
\end{equation}
where $E_0[\vec{A}^{\mathrm{ext}}]$ is the vacuum energy in
presence of the external background field. Thus we see that $
\Gamma[\vec{A}^{\mathrm{ext}}]$ is the lattice gauge-invariant
effective action for the static background field
$\vec{A}^{\mathrm{ext}}$. It is worthwhile to stress that our
lattice effective action is defined by means of the lattice
Schr\"odinger functional Eq.~(\ref{Zetalatt}) with the same
boundary fields at $x_4 = 0$ and $x_4 = T$. As a consequence we
have~\cite{Cea:1997ff}:
\begin{equation}
\label{Zetal} {\mathcal{Z}}[U^{\mathrm{ext}}] = \int
{\mathcal{D}}U \; e^{-S_W} \;,
\end{equation}
where $S_W$ is the familiar Wilson action and the functional
integral is defined over a four-dimensional hypertorus with the
``cold-wall'' at $x_4 = 0$:
\begin{equation}
\label{coldwall}
U_\mu(x)|_{x_4=0} = U^{\mathrm{ext}}_\mu  \,.
\end{equation}
In previous studies~\cite{Cea:1996sw,Cea:1998iy,Cea:1999gn} 
we investigated the vacuum
dynamics of the SU(2) lattice gauge theory by using an external
constant Abelian chromagnetic field. In that case the relevant
quantity is the density of effective action:
\begin{equation}
\label{density}
\varepsilon[\vec{A}^{\mathrm{ext}}] = -\frac{1}{\Omega} \ln \left[
\frac{{\mathcal{Z}}[A^{\mathrm{ext}}]}{{\mathcal{Z}}[0]} \right]
\,,
\end{equation}
where $\Omega = V \cdot T$, $V$ being the spatial volume. \\
The aim of the present paper is twofold. First, we extend the
study of the effective action with an external Abelian
chromomagnetic field to the lattice SU(3) gauge theory at zero
temperature. Second, we consider both SU(2) and SU(3)  gauge
systems at finite temperature. \\
At finite temperature the relevant quantity is the partition
function :
\begin{equation}
\label{thermal}
\text{Tr}\left[ e^{-\beta_T H} \right] = \int
\mathcal{D}\vec{A} \, \langle \vec{A} \left |
e^{-\beta_T H} \mathcal{P} \right| \vec{A} \rangle \,,
\end{equation}
where $\beta_T$ is the inverse of the physical temperature. The
thermal partition function Eq.~(\ref{thermal}) can be written as a
functional integral~\cite{Gross:1981br}:
\begin{equation}
\label{tpf}
\text{Tr}\left[ e^{-\beta_T H} \right] =
\int_{A_\mu(\beta_T,\vec{x})=A_\mu(0,\vec{x})}
 \mathcal{D}A_\mu(t,\vec{x})  \,
e^{-\int^{\beta_T}_0 dx_4 \, \int d^3 \vec{x}
\mathcal{L}_{Y-M}(\vec{x},x_4)} \,.
\end{equation}
The lattice implementation of Eq.~(\ref{tpf}) is straightforward.
We have:
\begin{equation}
\label{tpflattice}
\text{Tr}\left[ e^{-\beta_T H} \right] =
\int_{U_\mu(\beta_T,\vec{x})=U_\mu(0,\vec{x})=U_\mu(\vec{x})}
 \mathcal{D}U_\mu(x_4,\vec{x})  \,
e^{-S_W} \,.
\end{equation}
Note that, by comparing  Eq.~(\ref{tpflattice}) with
Eqs.~(\ref{Zetal}), we have:
\begin{equation}
\label{trace}
\text{Tr}\left[ e^{-\beta_T H} \right] = \int
\mathcal{D}U_\mu(\vec{x})  \, \mathcal{Z}[U_\mu(\vec{x})] \,,
\end{equation}
where $\mathcal{Z}[U_\mu(\vec{x})]$ is the Schr\"odinger
functional Eq.~(\ref{Zetal}) defined on a lattice with
$L_4=\beta_T$, and ``external'' links $U_\mu(\vec{x})$ at $x_4=0$.
\\
We are interested in the thermal partition function in presence of
a given static background field
$\vec{A}^{\mathrm{ext}}(\vec{x})$. In the continuum this
can be obtained by splitting the gauge field into the background
field $\vec{A}^{\mathrm{ext}}(\vec{x})$ and the
fluctuating fields $\vec{\eta}(\vec{x})$. So that we could write
formally:
\begin{equation}
\label{ZetaT}
\mathcal{Z}_T[\vec{A}^{\mathrm{ext}}] = \int \mathcal{D}
\vec{\eta} \, \langle \vec{A}^{\mathrm{ext}}, \vec{\eta}
\left| e^{-\beta_T H} \mathcal{P} \right|
\vec{A}^{\mathrm{ext}}, \vec{\eta} \rangle \,.
\end{equation}
The lattice implementation of Eq.~(\ref{ZetaT}) can be obtained
from Eq.~(\ref{tpflattice}) if we write
\begin{equation}
\label{ukbetat}
U_k(\beta_T,\vec{x})=U_k(0,\vec{x})=U^{\text{ext}}_k(\vec{x})
\widetilde{U}_k(\vec{x}) \,,
\end{equation}
where $U^{\text{ext}}_k(\vec{x})$ is the lattice version of the
external continuum field $\vec{A}^{\mathrm{ext}}$ and the
$\widetilde{U}_k(\vec{x})$'s are the fluctuating links. Thus we
get:
\begin{equation}
\label{zzz}
\mathcal{Z}_T[\vec{A}^{\mathrm{ext}}] = \int_{x_4=0}
\mathcal{D}\widetilde{U}_k(\vec{x}) \, \mathcal{D}{U}_4(\vec{x})
\, \mathcal{Z}[U_k^{\text{ext}}(\vec{x}),\widetilde{U}_k(\vec{x})]
\,.
\end{equation}
Note that in Eq.~(\ref{zzz}) only the spatial links belonging to
the hyperplane $x_4=0$ are written as the product of the external
link $U^{\text{ext}}_k(\vec{x})$ and the fluctuating links
$\widetilde{U}_k(\vec{x})$. The temporal links
$U_4(x_4=0,\vec{x})$ are left freely fluctuating. It follows that
the temporal links $U_4(x)$ satisfy the usual periodic boundary
conditions. We stress that the periodic boundary conditions in the
temporal direction are crucial to retain the physical
interpretation that the functional
$\mathcal{Z}_T[\vec{A}^{\text{ext}}]$ is a thermal
partition function. In the following the spatial links belonging
to the time-slice $x_4=0$ will be called ``frozen links'', while
the remainder will be the ``dynamical links''. From the physical
point of view we are considering the gauge system at finite
temperature in interaction with a fixed external background field.
As a consequence, in the Wilson action $S_W$ we keep only the
plaquettes built up with the dynamical links or with dynamical and
frozen links. With these limitations it is easy to see that in
Eq.~(\ref{zzz}) we have:
\begin{equation}
\label{ZetaText}
\mathcal{Z} \left[ U_k^{\text{ext}}(\vec{x}),
\widetilde{U}_k(\vec{x}) \right] = \mathcal{Z}
\left[U_k^{\text{ext}}(\vec{x}) \right] \,.
\end{equation}
Indeed, let us consider an arbitrary frozen link
$U^{\text{ext}}_k(\vec{x}) \widetilde{U}_k^{\text{ext}}(\vec{x})$.
This link enters in the modified Wilson action by means of the
plaquette:
\begin{equation}
\label{plaquette}
P_{k4}(x_4=0,\vec{x}) = \text{Tr} \left\{
U^{\text{ext}}_k(\vec{x})  \widetilde{U}_k^{\text{ext}}(\vec{x})
U_4(0,\vec{x}+\hat{k}) U^\dagger_k(1,\vec{x}+\hat{k})
U^\dagger_4(0,\vec{x}) \right\} \,.
\end{equation}
Now we observe that the link $U_4(0,\vec{x}+\hat{k})$ in
Eq.~(\ref{plaquette}) is dynamical, i.e. we are integrating over
it. So that, by using the invariance of the Haar measure we
obtain:
\begin{equation}
\label{plaqnew}
P_{k4}(x_4=0,\vec{x}) = \text{Tr} \left\{
U^{\text{ext}}_k(\vec{x}) U_4(0,\vec{x}+\hat{k})
U^\dagger_k(1,\vec{x}+\hat{k}) U^\dagger_4(0,\vec{x}) \right\} \,.
\end{equation}
It is evident that Eq.~(\ref{plaqnew}) in turns implies
Eq.~(\ref{ZetaText}). Then, we see that in Eq.~(\ref{zzz}) the
integration over the fluctuating links $\widetilde{U}_k(\vec{x})$
gives an irrelevant multiplicative constant. So that we obtain:
\begin{equation}
\label{ZetaTnew}
\mathcal{Z}_T \left[ \vec{A}^{\text{ext}} \right] =
\int_{U_k(\beta_T,\vec{x})=U_k(0,\vec{x})=U^{\text{ext}}_k(\vec{x})}
\mathcal{D}U \, e^{-S_W}   \,,
\end{equation}
where the integrations are over the dynamical links with periodic
boundary conditions in the time direction. As concern the boundary
conditions at the spatial boundaries, we keep the fixed boundary
conditions $U_k(\vec{x},x_4)=U_{k}^{\text{ext}}(\vec{x})$ used in
the Schr\"odinger functional Eq.~(\ref{Zetal}). We stress that, if
we send the physical temperature to zero, then the thermal
functional Eq.~(\ref{ZetaTnew}) reduces to the zero-temperature
Schr\"odinger functional Eq.~(\ref{Zetalatt}) with the constraints
$U_k(x)|_{x_4=0} = U^{\mathrm{ext}}_k(\vec{x})$ instead of
Eq.~(\ref{coldwall}). In our previous
studies~\cite{Cea:1997ff,Cea:1996sw,Cea:1998iy,Cea:1999gn} 
we checked that in the
thermodynamic limit both conditions agree as concern the
zero-temperature effective action Eq.~(\ref{Gamma}). \\
The plan of the paper is as follows. In Section 2 we consider the
Abelian constant chromomagnetic background field for the pure
gauge SU(3) lattice theory at zero temperature. Section 3 is
devoted to the study of both SU(2) and SU(3) lattice theories
at finite temperature in presence of the external background
field. Finally our conclusions are drawn in Section 4.
%
\renewcommand{\thesection}{\normalsize{\arabic{section}.}}
\section{\normalsize{SU(3) IN A CONSTANT CHROMOMAGNETIC FIELD}}
\renewcommand{\thesection}{\arabic{section}}

We are interested in the case of a constant Abelian chromomagnetic
field which in the continuum reads:
\begin{equation}
\label{field}
\vec{A}^{\mathrm{ext}}_a(\vec{x}) =
\vec{A}^{\mathrm{ext}}(\vec{x}) \delta_{a,3} \,, \quad
\vec{A}^{\mathrm{ext}}_k(\vec{x}) =  \delta_{k,2} x_1 H \,.
\end{equation}
The lattice links corresponding to
$\vec{A}^{\mathrm{ext}}(\vec{x})$ can be evaluated from:
\begin{equation}
\label{links}
U_\mu = {\mathrm P} \exp\left\{ i g \int_0^1 dt \,
A_\mu(x+t\hat{\mu}) \right\}
\end{equation}
where ${\mathrm P}$ is the path-ordering operator, and $A_\mu =
A_\mu^a \frac {\lambda^a }{2 }$, the  $\lambda^a $'s being the
Gell-Mann matrices. From Eqs.~(\ref{field}) and~(\ref{links}) we
get:
\begin{eqnarray}
\label{t3links}
U^{\mathrm{ext}}_1(\vec{x}) = & U^{\mathrm{ext}}_3(\vec{x}) =
U^{\mathrm{ext}}_4(\vec{x}) = {\mathbf{1}} \nonumber \\ \nonumber
\\
U^{\mathrm{ext}}_2(\vec{x}) = &
\begin{bmatrix}
e^{i \frac {g H x_1} {2} } & 0 & 0 \\ 0 &  e^{- i \frac {g H x_1}
{2} } & 0
\\ 0 & 0 & 1
\end{bmatrix}
\; \, .
\end{eqnarray}
Our Schr\"odinger functional Eq.~(\ref{Zetal}) is defined on a
lattice with the hypertorus geometry,  for it is natural to
impose that:
\begin{equation}
\label{pbcs}
U_2(x_1,x_2,x_3,x_4) = U_2(x_1+L_1,x_2,x_3,x_4) \,,
\end{equation}
where $L_1$ is the lattice extension in the $x_1$ direction (in
lattice units). As a consequence the magnetic field $H$ turns out
to be quantized:
\begin{equation}
\label{quant}
\frac {g H}{2} = \frac{2 \pi}{L_1} n_{\mathrm{ext}} \,,
\end{equation}
with $n_{\mathrm{ext}}$ integer. \\
According to the discussion of the previous Section, in evaluating
the lattice functional integral Eq.~(\ref{Zetal}) we impose that
links belonging to the time slice $x_4=0$ are frozen to the
configuration Eq.~(\ref{t3links}). Moreover we impose also that
the links at the spatial boundaries are fixed according to
Eq.~(\ref{t3links}). In the continuum this last condition amounts
to the usual requirement that the fluctuations over the background
fields vanish at  infinity. \\
An alternative possibility is given by constraining the links
belonging to the the time slice $x_4=0$ and those at the spatial
boundaries to the condition
\begin{equation}
\label{newpbcs}
U_k(x) = U_k^{\mathrm{ext}}(\vec{x}) \,, \qquad   k \, = \, 1,2,3
\end{equation}
while the links $U_4(x)$  are unconstrained. Note that with the
condition Eq.~(\ref{newpbcs}) the time-like plaquettes nearest the
frozen hypersurface $x_4=0$ behave symmetrically in the update
procedure. Moreover, in this way the Schr\"odinger functional
Eq.~(\ref{Zetal}) is the zero-temperature limit of the thermal
partition functional Eq.~(\ref{ZetaTnew}). In our previous
studies~\cite{Cea:1996sw,Cea:1998iy,Cea:1999gn} 
we checked that in the thermodynamic limit
both conditions  agree as the effective action is concerned. \\
Our numerical simulations at zero temperature have been done on a
lattice of size $L_1 L_2 L_3 L_4$ with $L_1=L_4=32$, while the
transverse size  $L_1 =L_2=L_\perp$ has been varied from
$L_\perp=8$ up to  $L_\perp=32$. To avoid the problem
 of computing a partition function which is the
exponential of an extensive quantity we
consider the derivative of the density of effective action
$\varepsilon[\vec{A}^{\mathrm{ext}}]$ Eq.~(\ref{density}) with
respect to $\beta$ by taking $n_{\mathrm{ext}}$ (i.e. $gH$) fixed.
Indeed, we have
\begin{eqnarray}
\label{epsilonprimo}
\lefteqn{ \varepsilon^\prime[\vec{A}^{\mathrm{ext}}]  =
\frac{\partial \varepsilon[\vec{A}^{\mathrm{ext}}]}{\partial
\beta}
 = -\frac{1}{\Omega} \left[ \frac{1}{{\mathcal{Z}}[U^{\mathrm{ext}}]}
\frac{\partial {\mathcal{Z}}[U^{\mathrm{ext}}]}{\partial \beta}
\right. }
 \nonumber \\ & &
  \left. - \frac{1}{{\mathcal{Z}}[0]}
\frac{\partial {\mathcal{Z}}[0]}{\partial \beta} \right] =
\left\langle \frac{1}{\Omega} \sum_{x, \mu> \nu} \frac{1}{3} \,
{\mathrm{Re} \, \mathrm{Tr}} U_{\mu \nu}(x) \right\rangle_0
\nonumber
\\
& & - \left\langle \frac{1}{\Omega} \sum_{x, \mu> \nu} \frac{1}{3}
\, {\mathrm{Re} \, \mathrm{Tr}} U_{\mu \nu}(x)
\right\rangle_{\vec{A}^{\mathrm{ext}}}\,,
\end{eqnarray}
where the subscripts on the average indicate the value of the
external links at the boundaries, and the $ U_{\mu \nu}(x)$'s are
the plaquettes in the $(\mu, \nu)$ plane. Actually, the
contributions to $\varepsilon^\prime[\vec{A}^{\mathrm{ext}}]$ due
to the frozen time-slice at $x_4=0$ and the fixed links at the
spatial boundaries must be subtracted. Accordingly, we define the
derivative of the internal energy density:
\begin{eqnarray}
\label{deriv}
\lefteqn{
\varepsilon^{\prime}_{\mathrm{int}}[\vec{A}^{\mathrm{ext}}] =
\left \langle \frac{1}{\Omega_{\mathrm{int}}} \sum_{x \in
\tilde{\Lambda},\mu > \lambda} \frac{1}{3} \,  {\mathrm{Re} \,
\mathrm{Tr}} U_{\mu\nu}(x) \right\rangle_0} \nonumber \\ & & -
\left\langle \frac{1}{\Omega_{\mathrm{int}}} \sum_{x \in
\tilde{\Lambda},\mu > \lambda} \frac{1}{3} \, {\mathrm{Re} \,
\mathrm{Tr}} U_{\mu\nu}(x) \right\rangle_{\vec{A}^{\mathrm{ext}}}
\,,
\end{eqnarray}
where $\tilde{\Lambda}$ is the ensemble of the internal lattice
sites which occupy the volume $\Omega_{\mathrm{int}}$. \\
To implement the constraint at the boundaries in the numerical
simulations we update only the internal links, i.e. the links
$U_\mu(x)$ with $x \in \tilde{\Lambda}$. We use the over-relaxed
heat-bath algorithm to update the gauge configurations.
Simulations have been performed by means of the APE100
computer. Since we are measuring a local quantity such as the
plaquette, a low statistics (from 1000 up to 5000 configurations)
is required in order to get a good estimate of
$\varepsilon^{\prime}_{\mathrm{int}}$.

In Figure~1 we display the derivative of the energy density
normalized to the derivative of the external energy density:
\begin{equation}
\label{epsprimeext}
\varepsilon^{\prime}_{\mathrm{ext}} = \frac{2}{3} \, [1 - \cos(
\frac{g H}{2} )] = \frac{2}{3} \, [1 - \cos( \frac{2 \pi}{L_1}
n_{\mathrm{ext}})]
\end{equation}
versus $\beta$ for $L_1=L_4=32$ and $8 \le L_\perp \le 32$.
From Figure~1 we see that, as in the SU(2) gauge
theory~\cite{Cea:1996sw,Cea:1998iy,Cea:1999gn}, 
both the peak and the perturbative plateau
of $\varepsilon^{\prime}_{\mathrm{int}}$ decrease by increasing
$L_\perp$. In order to perform the thermodynamic limit we
introduce the scaling variable:
\begin{equation}
\label{x-scaling}
x = \frac{a_H}{L_{\mathrm{eff}}} \,,
\end{equation}
where
\begin{equation}
\label{mag-length}
a_H = \sqrt{ \frac{2 \pi}{g H} }
\end{equation}
is the magnetic length, and
\begin{equation}
\label{L-eff}
L_{\mathrm{eff}} = \Omega_{\mathrm{int}}^{1/4}
\end{equation}
is the lattice effective linear size. As in the SU(2) case~\cite{Cea:1999gn}
we try the scaling law:
\begin{equation}
\label{universal}
x^{-\alpha}
\frac{\varepsilon^{\prime}_{\mathrm{int}}
(\beta,n_{\mathrm{ext}},L_{\mathrm{eff}})}
{\varepsilon^{\prime}_{\mathrm{ext}}}  = \kappa(\beta) \,.
\end{equation}
Indeed, from Figure~2 we see that our numerical data can be
arranged on the scaling curve $ \kappa(\beta)$. It is remarkable
that the value of the exponent $\alpha = 1.5$ in
Eq.~(\ref{universal}) agrees with the one we found for the SU(2)
gauge theory~\cite{Cea:1996sw,Cea:1998iy,Cea:1999gn}. 
From Eq.~(\ref{universal}) we can
determine the infinite volume limit of the vacuum energy density
$\varepsilon_{\mathrm{int}}$. We have:
\begin{equation}
\label{infinite-vol}
\lim_{L_{\mathrm{eff}} \to \infty}
\varepsilon_{\mathrm{int}}(\beta,n_{\mathrm{ext}},L_{\mathrm{eff}})=
 \varepsilon^{\prime}_{\mathrm{ext}} \,
\int_0^{\beta} d{\beta^{\prime}} \, \kappa(\beta^{\prime})
\lim_{L_{\mathrm{eff}} \to \infty} \left(
\frac{a_H}{L_{\mathrm{eff}}} \right)^\alpha  = 0
\end{equation}
in the whole range of $\beta$. As a consequence, in the continuum
limit $(L_{\mathrm{eff}} \to \infty, \beta \to \infty)$ the SU(3)
vacuum screens the external chromomagnetic Abelian field in
accordance with the dual superconductivity
scenario~\cite{tHooft:1976eps,Mandelstam:1974pi}.
%
%
\renewcommand{\thesection}{\normalsize{\arabic{section}.}}
\section{\normalsize{FINITE TEMPERATURE BACKGROUND FIELD EFFECTIVE ACTION}}
\renewcommand{\thesection}{\arabic{section}}
Let us consider the gauge systems in an external chromomagnetic
Abelian field at finite temperature. According to the discussion
in Section 1, we are interested in the thermal partition function
$\mathcal{Z}_T \left[ \vec{A}^{\text{ext}} \right]$,
Eq.~(\ref{ZetaTnew}). On the lattice the physical temperature
$T_{physical}$ is given by:
\begin{equation}
\label{temperature}
\frac{1} {T_{physical}} \, = \, L_t
\end{equation}
where $L_t=L_4$ is the linear extension in the time direction. In
order to approximate the thermodynamic limit, the spatial
extension $L_s$ should respect the relation:
\begin{equation}
\label{thermodinamic}
 L_s  \, \gg \, L_t \, .
\end{equation}
To this end we perform our numerical simulation on $32^3 \times
L_t$ and  $64^3 \times L_t$ lattices by imposing:
\begin{equation}
\label{latticetemp}
\frac{L_t} {L_s} \, \leq \, 4 \, .
\end{equation}
At finite temperature the effective action is defined through the
free energy:
\begin{equation}
\label{freeenergy}
F[\vec{A}^{\mathrm{ext}}] = -\frac{1}{L_t} \ln \left\{
\frac{\mathcal{Z}_T \left[ \vec{A}^{\text{ext}}
\right]}{{\mathcal{Z}_T}(0)} \right\} \,.
\end{equation}
Obviously, in the case of constant external chromagnetic field the
relevant quantity is the density of effective action:
\begin{equation}
\label{freeenergydensity}
f[\vec{A}^{\mathrm{ext}}] \, = \,  \frac{1} {V} \,
F[\vec{A}^{\mathrm{ext}}] \, .
\end{equation}
As is well known, by increasing the temperature the pure gauge
system undergoes the deconfinement phase transition. In the case
of pure SU(N) gauge theories it is known that the expectation value of
the Polyakov loop in the time direction
\begin{equation}
\label{polyakov}
P = \frac{1}{V_s} \sum_{\vec{x}} \frac{1}{N} {\text{Tr}} \prod_{x_4=1}^{L_t} 
U_4(\vec{x},x_4)
\end{equation}
is the order parameter for the deconfinement phase transition. As a
preliminary step we look at the behavior of the temporal Polyakov
loop  versus the external applied field. We
start with the SU(2) gauge system at $\beta=2.5$ on $32^3 \times
5$ lattice at zero applied external field (i.e.
$n_{\mathrm{ext}}=0$) that is known to be in the deconfined phase
of finite temperature SU(2). If the external field strength is
increased the expectation value of the Polyakov loop is driven
towards the value at zero temperature (see Fig.~3). 
It is worthwhile to stress that this last result is consistent with the dual
superconductor mechanism of confinement.
Similar behavior has been reported by Authors of
Refs.~\cite{Meisinger:1997jt,Ogilvie:1998my} within a different approach. 
If we now consider the SU(2) gauge system at zero
temperature in a constant Abelian chromomagnetic background field
of fixed strength ($n_{\mathrm{ext}}=1$) and increase the
temperature, we find that the perturbative tail of the
$\beta$-derivative of the free energy density
$f^{\prime}_{\mathrm{int}}(\beta,n_{\mathrm{ext}})/
\varepsilon^{\prime}_{\mathrm{ext}}$ increases with $1/L_t$ and
tends towards the ``classical'' value
$f^{\prime}_{\mathrm{int}}(\beta,n_{\mathrm{ext}})/
\varepsilon^{\prime}_{\mathrm{ext}} \simeq 1 $ (see Fig.~4).
Therefore we may  conclude that, as the temperature increases, 
there is no screening effect in the free energy density, confirming that
the zero-temperature screening of the external field is related to
the confinement. 

The knowledge of
$f^{\prime}_{\mathrm{int}}(\beta,n_{\mathrm{ext}})/
\varepsilon^{\prime}_{\mathrm{ext}}$ at finite temperature can be
used to estimate the deconfinement temperature $T_c$. In
Figure~5 we magnify the peak region for different values of $L_t$.
We see clearly that the pseudocritical coupling $\beta^*(L_t)$
depends on $L_t$. To determine the pseudocritical couplings we
parametrize $f^{\prime}_{\mathrm{int}}(\beta,L_t)$ near the peak
as:
\begin{equation}
\label{peak-form}
\frac{f^{\prime}_{\mathrm{int}}(\beta,L_t)}{\varepsilon^{\prime}_{\mathrm{ext}}}
= \frac{a_1(L_t)}{a_2(L_t) [\beta - \beta^*(L_t)]^2 +1} \,.
\end{equation}
We restrict the region near $\beta^*(L_t)$ until the fits
Eq.~(\ref{peak-form}) give a reduced $\chi^2/{\text d.o.f.} \sim 1$.
\\
Having determined $\beta^*(L_t)$ we estimate the deconfinement
temperature as:
\begin{equation}
\label{Tc}
\frac{T_c}{\Lambda_{\mathrm{latt}}} = \frac{1}{L_t}
\frac{1}{f(\beta^*(L_t))} \,,
\end{equation}
where
\begin{equation}
\label{af}
f(\beta) = \left( \frac{\beta}{2  N b_0} \right)^{51/121} \, 
\exp \left( -\beta \frac{1}{4 N b_0} \right) \,,
\end{equation}
where $b_0=(11 N)/(48 \pi^2)$ and $N$ is the color number.
In Figure~6 we display $T_c/\Lambda_{\mathrm{latt}}$ for different
temperatures. Following Ref.~\cite{Fingberg:1993ju} we 
linearly extrapolate to the continuum our data for
$T_c/\Lambda_{\mathrm{latt}}$. In this way we obtain the following estimate
of the critical temperature in the continuum limit
\begin{equation}
\label{ourcritical}
 \frac{T_c} {\Lambda_{\mathrm{latt}}} \, = \, 28.36 \,\pm \, 1.38 \, .
\end{equation}
Equation~(\ref{ourcritical}) is to be compared with the continuum
limit of the critical temperature available in the
literature~\cite{Fingberg:1993ju}
\begin{equation}
\label{critical}
 \frac{T_c} {\Lambda_{\mathrm{latt}}} \, = \, 24.38 \,\pm \, 2.18 \, .
\end{equation}
Our result Eq.~(\ref{ourcritical}) agrees, within two standard deviations,  with the
result given in  Eq.~(\ref{critical}). Note, however, that the small discrepancy 
could be a true dynamical effect due to the external chromomagnetic field.
Indeed, the behaviour of the Polyakov loop versus the external
background field displayed in Fig.~3 seems to suggest that the
critical temperature does depend on the applied background field.
For dimensional reasons one expects that:
\begin{equation}
\label{magcrit}
 T_c^2 \, \, \sim \,  \, H \, .
\end{equation}
To check the expected behaviour  Eq.~(\ref{magcrit}) we need to
vary the external chromomagnetic field. We plan to do this study
in a future work.\\
For the time being, let us turn to the SU(3) pure
gauge lattice theory at finite temperature.
In this paper we limit ourselves to present
our determination of the critical deconfinement temperature.
In Figure~7 we display the peak region of the internal free energy density 
for different values of $L_t$, together with the fits
Eq.~(\ref{peak-form}). Having determined the pseudocritical
couplings, the deconfinement temperature can be obtained from
Eq.(\ref{Tc}).
Performing the linear extrapolation to the continuum limit we get:
\begin{equation}
\label{ourcriticalsu3}
 \frac{T_c} {\Lambda_{\mathrm{latt}}} \, = \, 20.86 \,\pm \, 3.02 \, .
\end{equation}
Our estimate Eq.~(\ref{ourcriticalsu3}) is in fair agreement with
the continuum limit of the SU(3) critical temperature available in the
literature~\cite{Fingberg:1993ju}
\begin{equation}
\label{criticalsu3}
 \frac{T_c} {\Lambda_{\mathrm{latt}}} \, = \, 29.67 \,\pm \, 5.47 \,.
\end{equation}
However, we note once more that our data do not exclude a
dependence of the critical temperature on the  external
chromomagnetic field.
\renewcommand{\thesection}{\normalsize{\arabic{section}.}}
\section{\normalsize{CONCLUSIONS}}
\renewcommand{\thesection}{\arabic{section}}
We have studied the non-perturbative dynamics of the vacuum of
SU(2) and SU(3) lattice gauge theories by means of a
gauge-invariant effective action defined using the lattice
Schr\"odinger functional.

At zero temperature our numerical results indicate that even for
the more interesting case of the SU(3) theory, in the continuum
limit $L_{\mathrm{eff}} \to \infty$, $\beta \to \infty$ we have:
\begin{equation}
\label{climit}
\varepsilon[ \vec{A}^{\mathrm{ext}} ] = 0 \,,
\end{equation}
so that the SU(3) vacuum screens completely the external
chromomagnetic Abelian field. In other words, the continuum vacuum
behaves as an Abelian magnetic condensate medium in accordance
with the dual superconductivity scenario.

The intimate connection between the screening of the external
background field and the confinement is corroborated by the finite
temperature results. Indeed our numerical data show that the
zero-temperature screening of the external field is removed by
increasing the temperature. Moreover, at finite temperature it
seems that confinement is restored by increasing the strength of
the external applied field.

At finite temperature we find that the $\beta$-derivative of the
free energy density behaves like a specific heat. From the peak
position of the $\beta$-derivative of the free energy density we
obtained an estimate of the critical temperature
$T_c/\Lambda_{\mathrm{latt}}$ that extrapolates in the continuum
limit to  values in fair agreement with  previous determinations
in the literature. Moreover, our data are suggestive of a
non-trivial dependence of the deconfinement critical temperature
on the applied external chromagnetic field. We deserve to
a future work the investigation of this interesting possibility.
%
%
%

%
%
\vfill
\newpage

\begin{figure}[H]
\label{Fig1}
\begin{center}
\includegraphics[clip,width=1.00\textwidth]{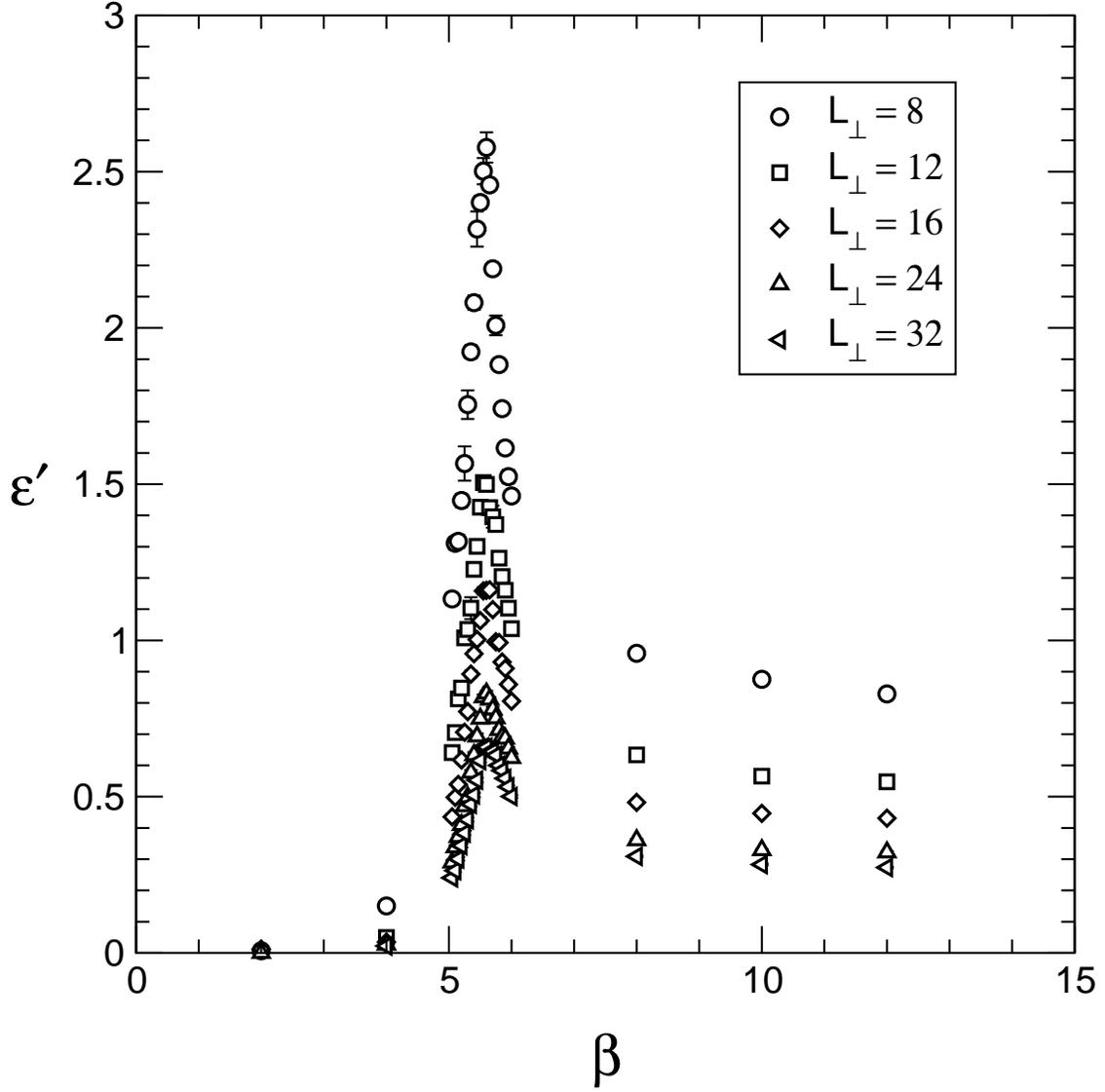}
\caption{The $\beta$-derivative of the internal energy density
Eq.~(\ref{deriv}) versus $\beta$ for SU(3) on a 
$L \times L_\perp^2 \times L$ lattice. Plotted data refer to $L=32$ and
different values of the transverse lattice size $L_\perp$.}
\end{center}
\end{figure}
\begin{figure}[H]
\label{Fig2}
\begin{center}
\includegraphics[clip,width=1.00\textwidth]{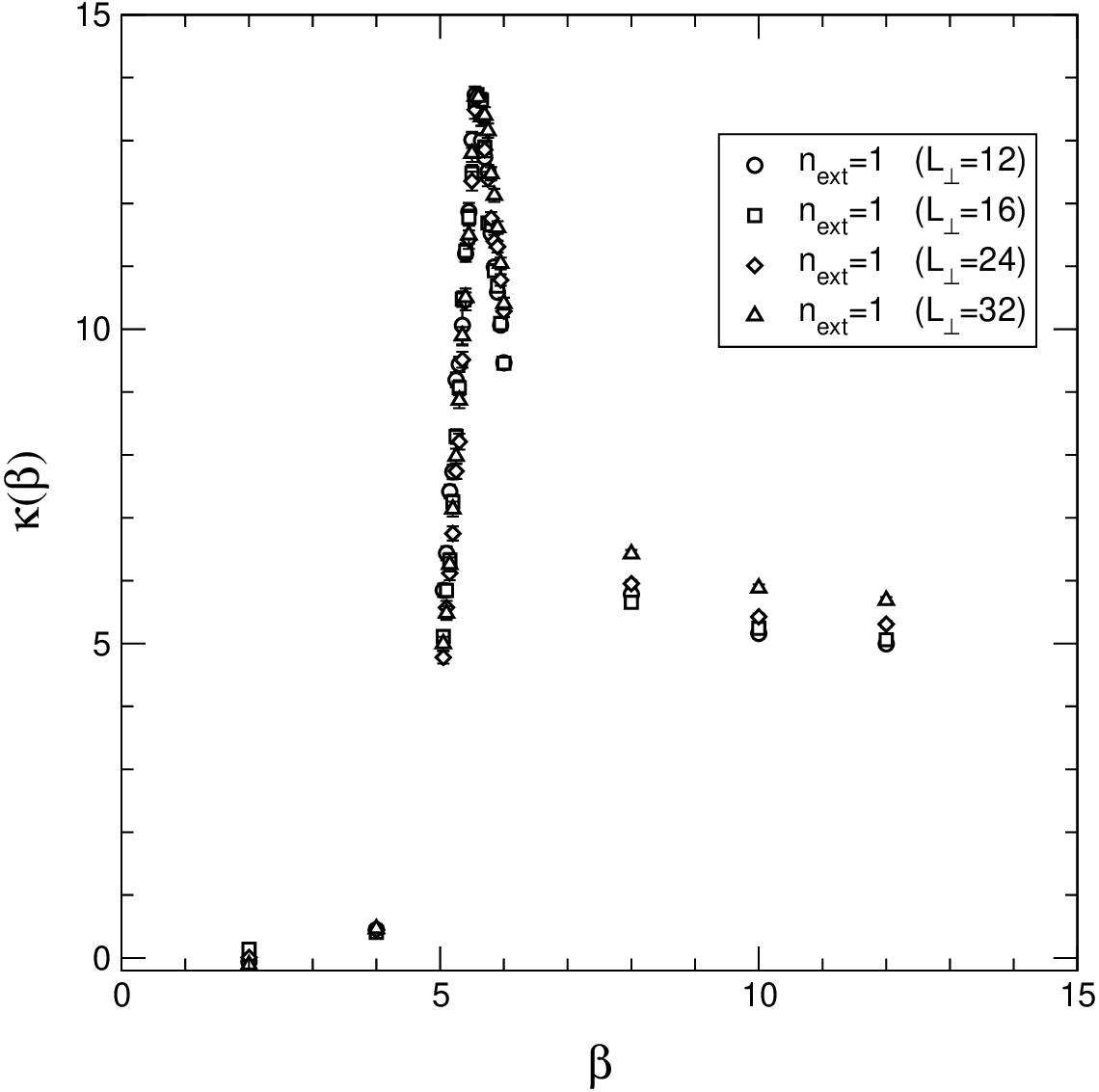}
\caption{The scaling curve for SU(3) obtained by re-scaling all lattice data
for
$\varepsilon^{\prime}_{\mathrm{int}}(\beta,n_{\mathrm{ext}},L_{\mathrm{eff}})$
according to Eq.~(\ref{universal}). }
\end{center}
\end{figure}
\begin{figure}[H]
\label{Fig3}
\begin{center}
\includegraphics[clip,width=1.00\textwidth]{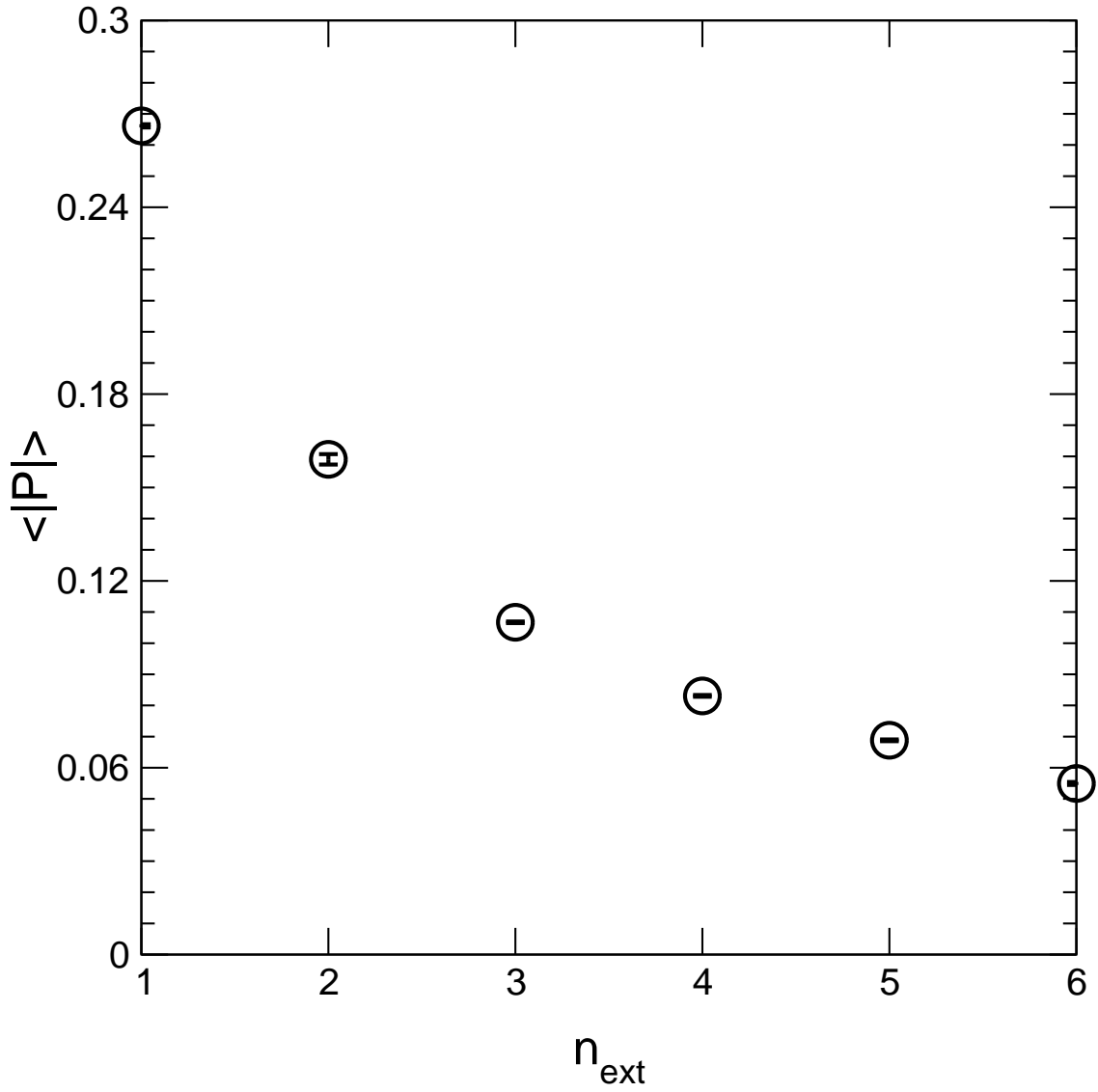}
\caption{The absolute value of the Polyakov loop versus $n_{\mathrm{ext}}$, for SU(2), on a
$32^3\times5$ lattice at $\beta=2.5$.}
\end{center}
\end{figure}
\begin{figure}[H]
\label{Fig4}
\begin{center}
\includegraphics[clip,width=1.00\textwidth]{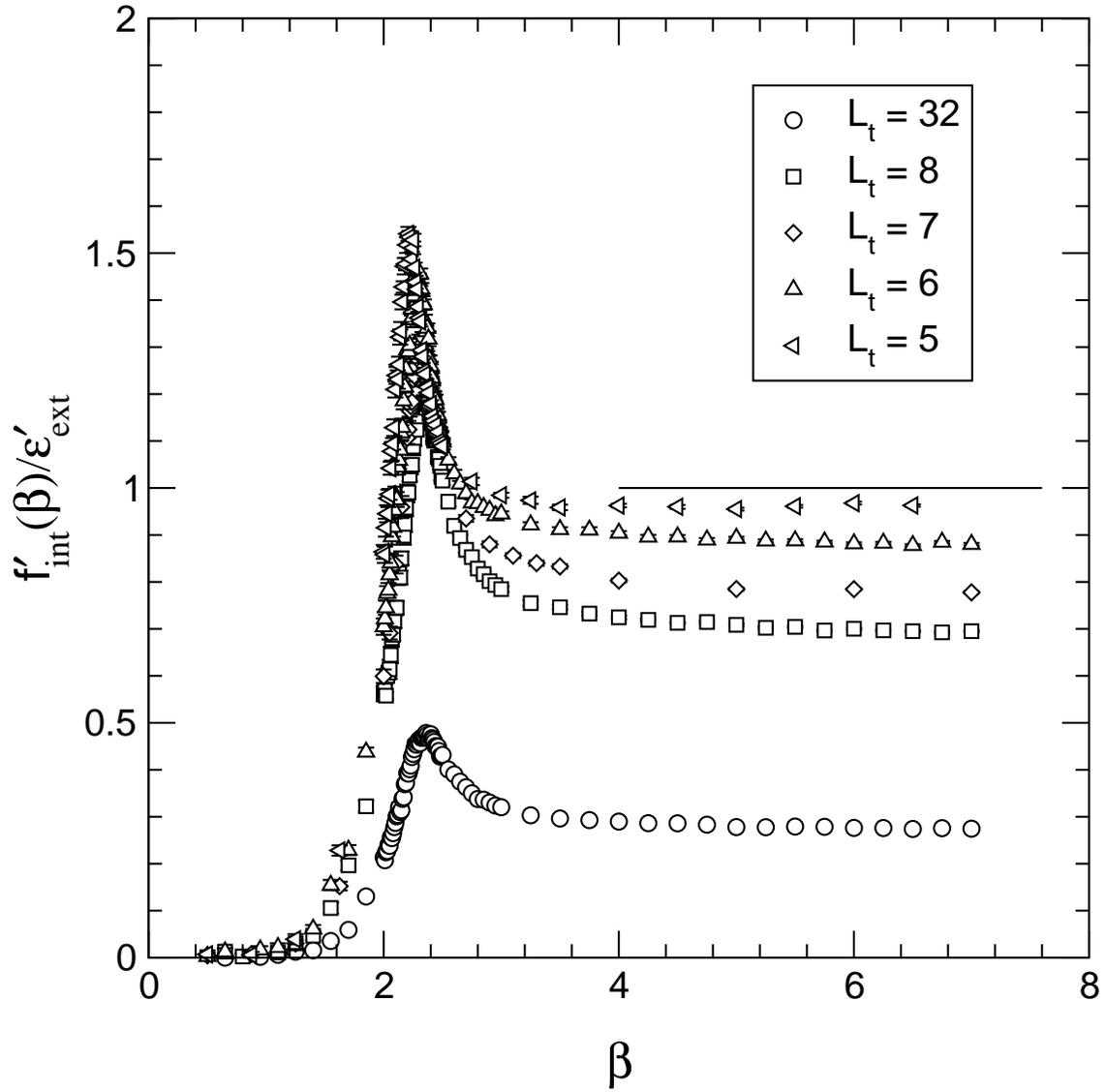}
\caption{The $\beta$-derivative of the free energy density
Eq.~(\ref{freeenergydensity}) for SU(2) 
on  $L_s^3 \times L_t$  lattices, with $L_s=32$.}
\end{center}
\end{figure}
\begin{figure}[H]
\label{Fig5}
\begin{center}
\includegraphics[clip,width=1.00\textwidth]{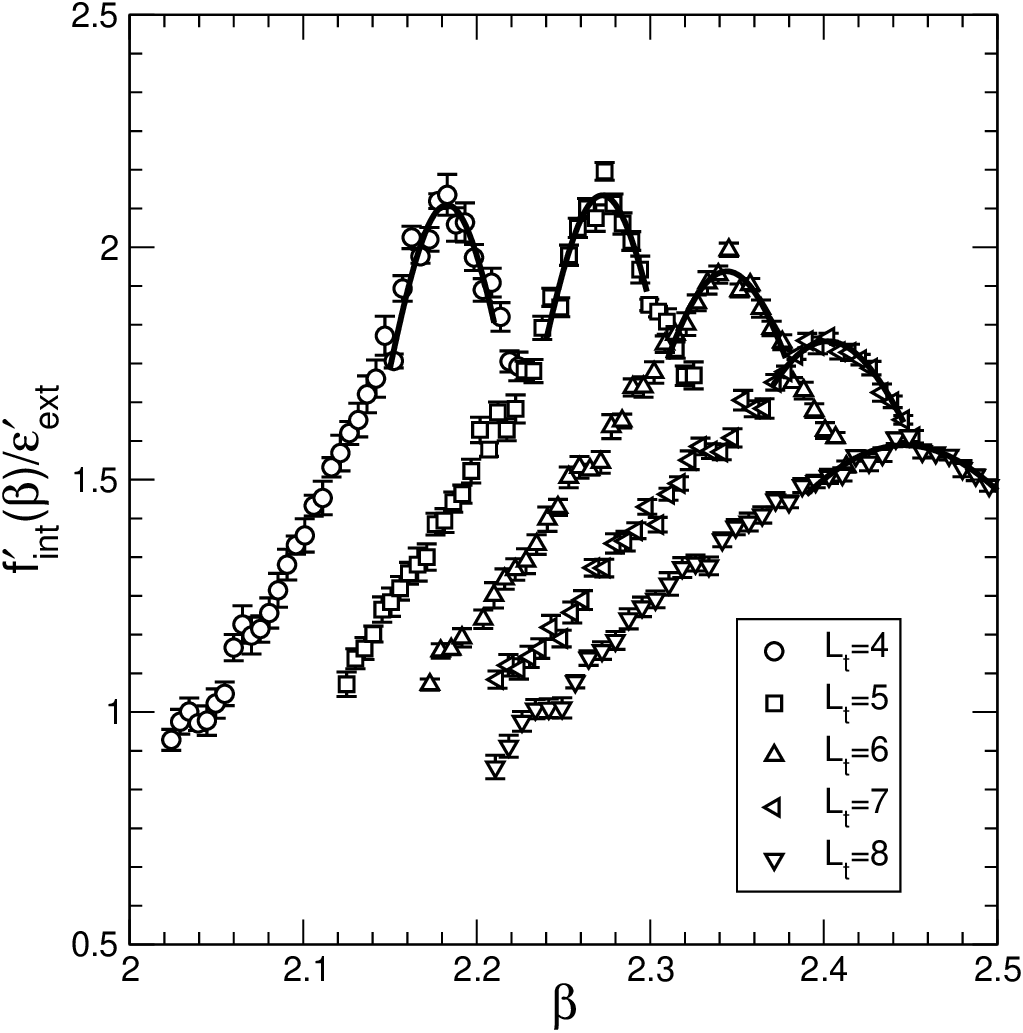}
\caption{The peak region for the 
$\beta$-derivative of the free energy density
Eq.~(\ref{freeenergydensity}) for SU(2).
The data are taken on $L_s^3 \times L_t$ lattices, with $L_s=64$.
Solid lines are the best fits to the data using Eq.~(\ref{peak-form}).}
\end{center}
\end{figure}
\begin{figure}[H]
\label{Fig6}
\begin{center}
\includegraphics[clip,width=1.00\textwidth]{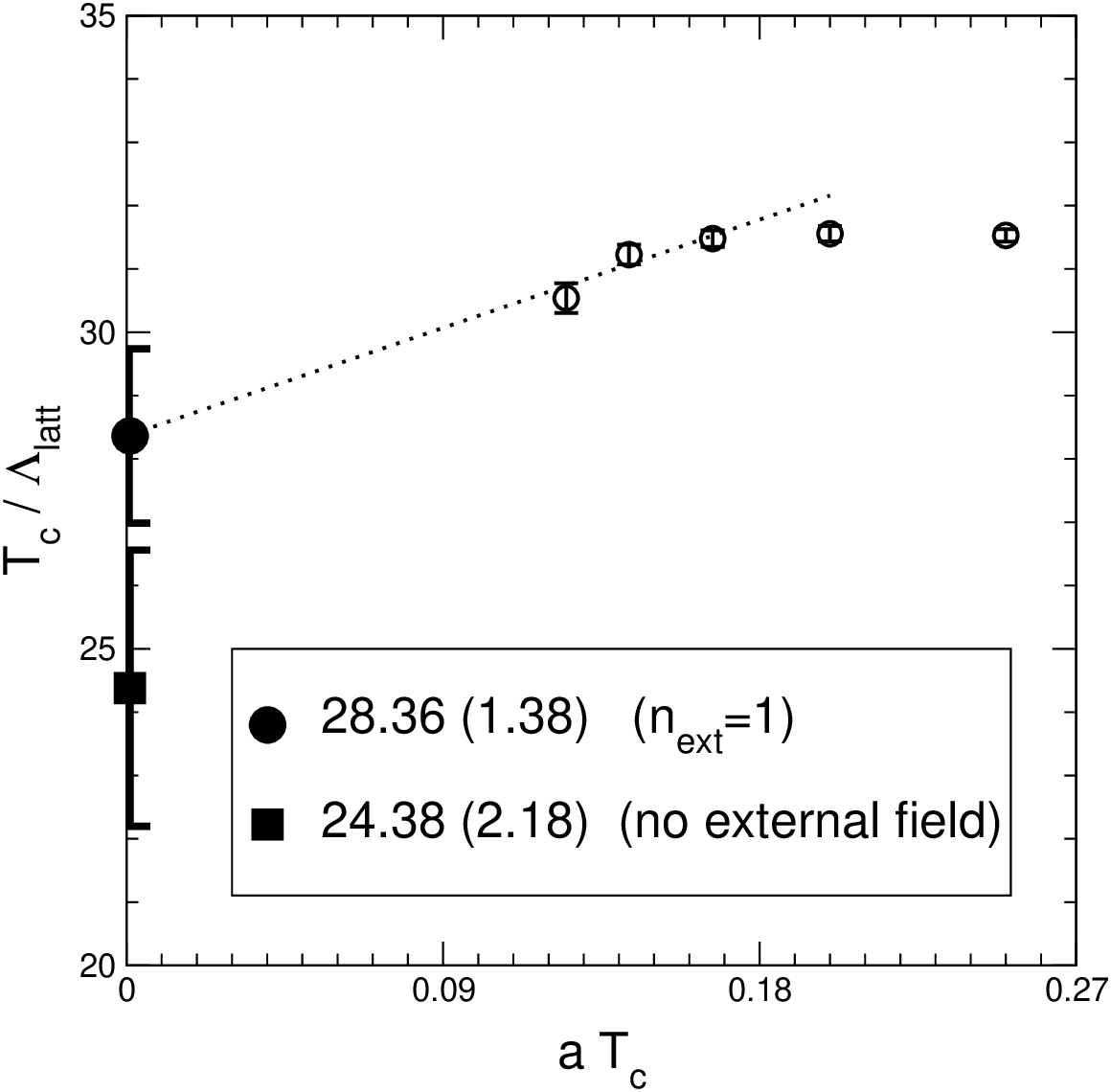}
\caption{Our SU(2) lattice data for
$T_c/\Lambda_{\mathrm{latt}}$ versus the temperature (circles).
The full square is the continuum extrapolation of
Ref.~\protect\cite{Fingberg:1993ju} (see Eq.~(\ref{critical})). The
dotted line is a linear fit to our data. }
\end{center}
\end{figure}
\begin{figure}[H]
\label{Fig7}
\begin{center}
\includegraphics[clip,width=1.00\textwidth]{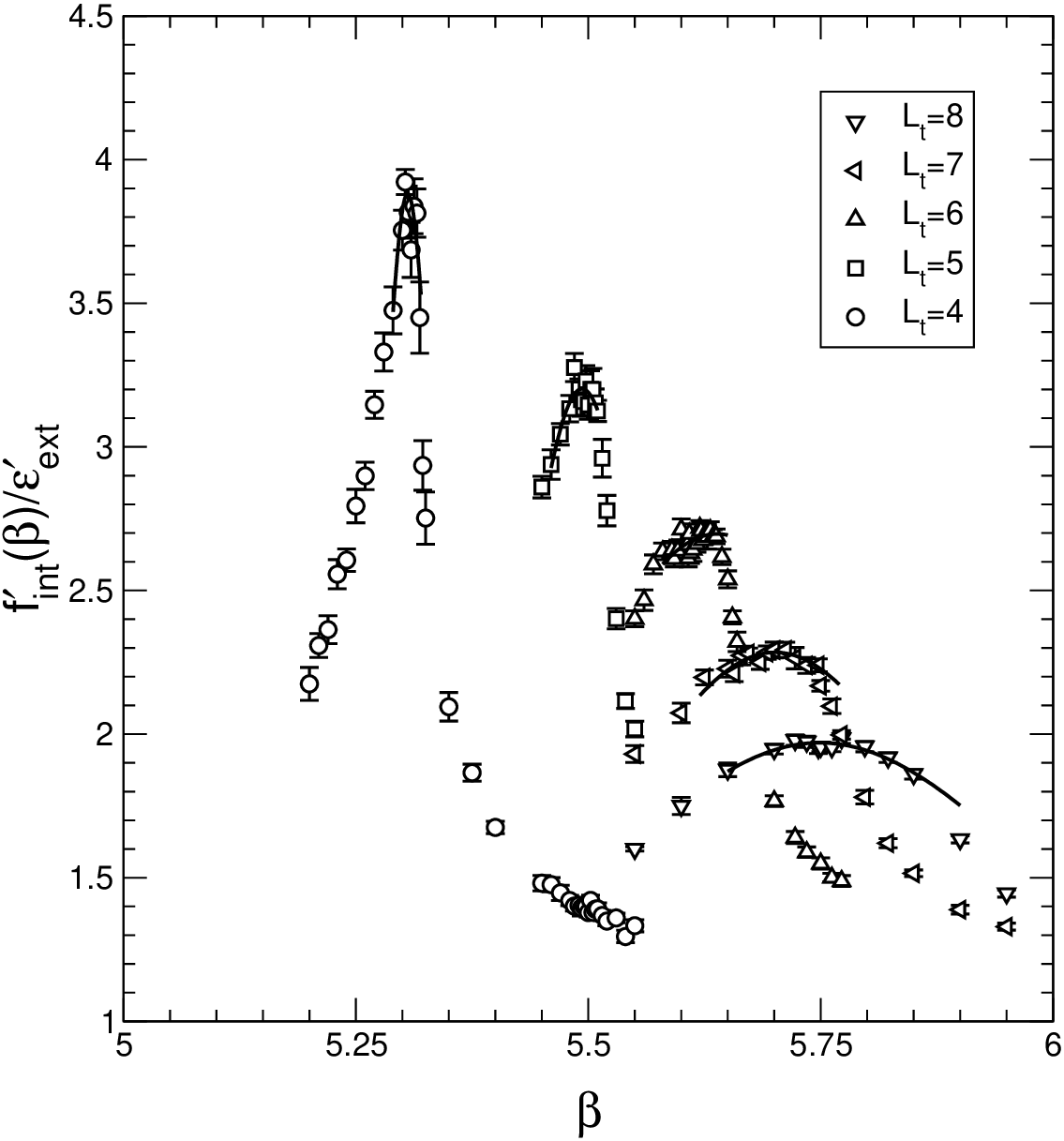}
\caption{The peak region for the 
$\beta$-derivative of the free energy density
Eq.~(\ref{freeenergydensity}) for SU(3).
The data are taken on $L_s^3 \times L_t$ lattices, with $L_s=64$.
Solid lines are the best fits to the data using Eq.~(\ref{peak-form}).}
\end{center}
\end{figure}
\begin{figure}[H]
\label{Fig8}
\begin{center}
\includegraphics[clip,width=1.00\textwidth]{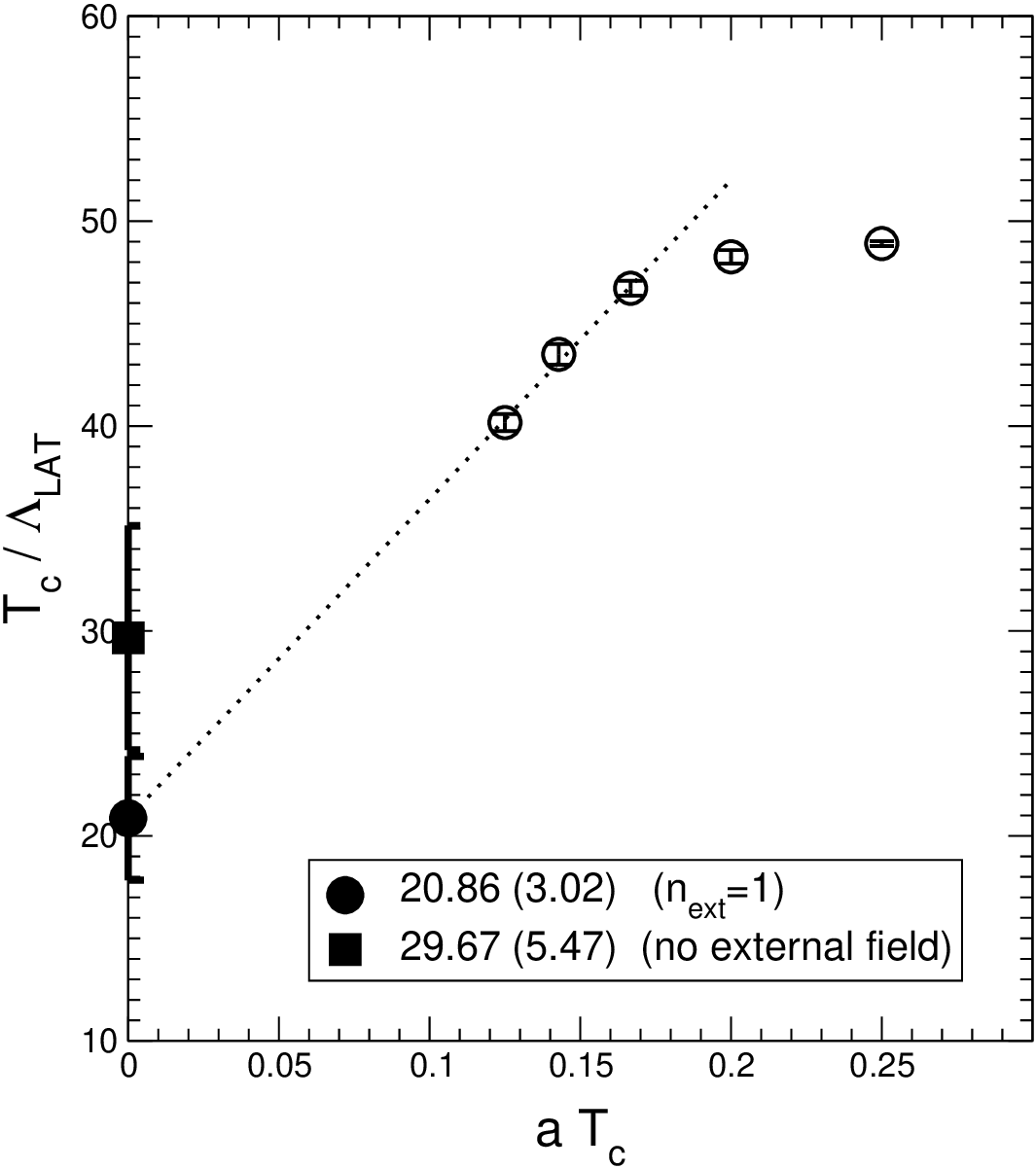}
\caption{Our SU(3) lattice data for
$T_c/\Lambda_{\mathrm{latt}}$ versus the temperature (circles).
The full square is the continuum extrapolation of
Ref.~\protect\cite{Fingberg:1993ju} (see Eq.~(\ref{criticalsu3})).
The dotted line is a linear fit to our data. }
\end{center}
\end{figure}
\end{document}